


\input phyzzx.tex

\tolerance=3000

\def\complex{{\bf C}}
\def\integer{{\bf Z}}
\def\natural{{\bf N}}
\def\positive{{\bf Z}^+}
\def\ie{{\it i.e.}\ }
\def\half{{1 \over 2}}
\def\la{\langle}
\def\ra{\rangle}
\def\tensor{\otimes}
\def\dtensor{\dot{\otimes}}
\def\hthree{h_3}
\def\hfour{h_4}
\def\sltwo{sl_2}

\def\hqthree#1{U_{#1} (\hthree)}
\def\hqfour#1{U_{#1} (\hfour)}
\def\slqtwo#1{U_{#1} (\sltwo)}
\def\ideal#1{\la #1 \ra}
\def\free#1#2{#1 [ \negthinspace [ #2 ] \negthinspace ]}
\def\cfree#1{ \free{\complex}{#1} }
\def\setof#1{ \lbrace #1 \rbrace}
\def\spacedcdot{\thinspace\cdot\thinspace}
\def\comma{,\enspace}

\def\Remark{{\sl Remark:\ }}

%

\def\ref{\REF}
\def\cite#1{${}^{#1}$}

\def\journal#1{{\it #1}}
\def\plb{\journal{Phys.\ Lett.\ {\bf B}}}
\def\npb{\journal{Nucl.\ Phys.\ {\bf B}}}
\def\cmp{\journal{Comm.\ Math.\ Phys.\ }}
\def\ijmp{\journal{Int.\ J.\ Mod.\ Phys.\ {\bf A}}}
\def\jpa{\journal{J.\ Phys.\ A}}
\def\lmp{\journal{Lett.\ Math.\ Phys.\ }}
\def\mpla{\journal{Mod.\ Phys.\ Lett.\ {\bf A}}}
\def\jmp{\journal{J.\ Math.\ Phys.\ }}
\def\pubrims{\journal{Publ.\ R.I.M.S.\ }}
\def\progInMath{\journal{Prog.\ in Math.\ }}
\def\leningrad{\journal{Leningrad Math.\ J.\ }}
\def\uspekhi{\journal{Uspekhi Mat.\  Nauk.\ }}

\ref\jimbo{M. Jimbo, {\it `A q-difference analogue of U(g) and the
	      Yang-Baxter equation'}, \lmp{\bf 10}, 63 (1985)}
\ref\drinfeld{V.G. Drinfeld, {\it `Quantum groups'},
	      in {\sl Proceedings of the I.C.M. at Berkeley 1986},
	      ed. A. Gleason (AMS, 1987), pp. 798-820}
\ref\frtworon{L. D. Faddeev, N. Yu. Reshetikhin and L. A. Takhtajan,
	      {\it `Quantisation of Lie groups and Lie algebras'},
	      \leningrad {\bf 1}, 193 (1990) \hfill \break
	S. L. Woronowicz, {\it `Compact matrix pseudogroups'},
	      \cmp {\bf 111}, 613 (1987)}
\ref\reviews{S. Majid, {\it `Quasitriangular Hopf algebras and Yang-Baxter
	       equations'}, \ijmp{\bf 5}, 1 (1990)\hfill\break
     M. Jimbo, {\it `Topics from representations of $U_q(g)$ -
	     an introductory guide for physicists'}, Nankai Lectures 1991,
		in {\sl Quantum Group and Quantum Integrable systems},
		eds. M. L. Ge, (World Scientific, 1992) \hfill \break
     T. Tjin, {\it `An introduction to quantised Lie groups and algebras'},
	    Amsterdam ITF preprint (1991), {\tt hep-th/9111043},
	    to be published in Int. J. Mod. Phys. A \hfill \break
     P. Aschieri and L. Castellani, {\it `An introduction to
	    non-commutative differential geometry on quantum groups'},
	    CERN-TH.6565/92 and Torino DFTT-22/92 preprint,
	    {\tt hep-th/9207084} (1992)}
\ref\sklyanin{E. K. Sklyanin, {\it `On an algebra generated by quadratic
        relations'}, \uspekhi {\bf 40}, 214 (1985)}
\ref\woronoSU{S. L. Woronowicz, {\it `Twisted $SU(2)$ group.
	an example of a non-commutative differential calculus'},
	\pubrims {\bf 23}, 117 (1987)}
\ref\witten{E. Witten, {\it`Gauge theories, vertex models
			and quantum groups'}, \npb {\bf 330}, 285 (1990)}
\ref\fairlie{D. Fairlie, {\it `Quantum deformation of $su(2)$'}, \jpa
					    {\bf 23}, L183 (1990)}
\ref\rosso{M. Rosso, {\it `Finite dimensional representations of the
quantum analogue of the enveloping algebra of complex simple Lie algebras'},
	\cmp {\bf 117}, 581 (1988)}
\ref\curtZach{T. L. Curtright and C. K. Zachos, {\it `Deforming maps
	for quantum algebras'}, \plb {\bf 243}, 237 (1990)}
%
%
\ref\macBieden{A. Macfarlane, \jpa {\bf 22}, 4581 (1989) \hfill \break
	       L. Biedenharn, \jpa {\bf 22}, L893 (1989)}
\ref\chaiKulLuk{M. Chaichian, P. Kulish and J. Lukierski,
	{\it `q-deformed Jacobi identity, q-oscillators and
	q-deformed infinite-dimensional algebras'}, \plb{\bf 237},
						 401 (1990) \hfill \break
	M. Chaichian and P. Kulish, {\it `Quantum superalgebras,
	q-oscillators and applications'}, preprint CERN-TH.5969/90 (1990)}
\ref\oscillref{P. Kulish and E. Damashinsky, {\it `On the q oscillator and
the quantum algebra $su_q(1,1)$'}, \jpa {\bf 23}, 981 (1990)}
\ref\yan{H. Yan, {\it `q-deformed oscillator algebra as a quantum group'}
		\jpa {\bf 23}, L1155 (1990)}
\ref\beijing{Z. Chan, W. Chen and H.-Y. Guo, \jpa {\bf 23},
					 4185 (1990) \hfill \break
	Z. Chan, W. Chen, H.-Y. Guo and H. Yang, \jpa {\bf 23}, 5371 (1990);
						       \hfill \break
	{\bf 24}, 1427 (1991); {\bf 24}, 5435 (1991)}
\ref\CGST{E. Celeghini, R. Giachetti, E. Sorace and M. Tarlini,
		     {\it `The quantum Heisenberg group $H(1)_q$'},
		     \jmp {\bf 32}, 1155 (1991) \hfill \break
	    C. G\'omez and G. Sierra, {\it `Quantum harmonic
		     oscillator algebra and link invariants'},
		     CSIC preprint, {\tt hep-th/9111005} (1991)}
%
%
\ref\fairlieGelfand{I. M. Gelfand and D. B. Fairlie,
	 {\it `The algebra of Weyl symmetrised polynomials
	 and its quantum extension'}, \cmp {\bf 136}, 487 (1991)}
\ref\woronoDG{S. L. Woronowicz, {\it `Differential calculus
     on compact matrix pseudogroups (quantum groups)'},
					  \cmp {\bf 122}, 125 (1989)}
\ref\manin{Yu. I. Manin, \cmp {\bf 123}, 163 (1989)}
\ref\wz{J. Wess and B. Zumino, {\it `Covariant differential calculus on
the quantum hyperplane'}, \npb\ (Proc. Suppl.) {\bf 18}, 302 (1990)}
\ref\zumino{B. Zumino, {\it `Deformation of the quantum mechanical
	      phase space with bosonic or fermionic coordinates'},
	      \mpla{\bf 6}, 1225 (1991)}
\ref\tensorAlgebra{J. E. Humphreys, {\sl `Introduction to Lie algebras and
representation theory'}, (Springer Verlag, 1972), chapter V}
\ref\castellani{L. Castellani, {\it `Bicovariant differential calculus
	      on the quantum D=2 Poincar\'e Group'},
	      {\tt hep-th/9201016}, \plb{\bf 279}, 291 (1992)}
\ref\conciniKac{C. De Concini and V. G. Kac, {\it `Representations
		of quantum groups at roots of 1'}, Colloque Dixmier,
		\progInMath {\bf 92}, pp. 471-506, (Birkh\"auser, 1990)}
\ref\cyclic{E. Date, M. Jimbo, K. Miki and T. Miwa,
{\it `Cyclic representations of $U_q(sl(n+1,\complex))$ at $q^N=1$'},
Kyoto preprint R.I.M.S.--703, (1990)}
\ref\sunGe{C.-P. Sun and M.-L. Ge, {\it `The cyclic boson operators
and new representations of the quantum algebra $sl_q(2)$
for q a root of unity'}, \jpa {\bf 24}, L969 (1991)}
\ref\filippov{A. T. Filippov, A. P. Isaev and A. B. Kurdikov,
{\it `Paragrassmann analysis and quantum groups'},
{\ tt hep-th/9204089}, \mpla{\bf 7}, 2129 (1992)}
\ref\celeghini{E. Celeghini, R. Giachetti, E. Sorace and M. Tarlini,
    {\it `Three-dimensional quantum groups from contractions of $SU(2)_q$'},
    \jmp {\bf 31}, 2548 (1990)}
\ref\kassel{C. Kassel, {\it `Cyclic homology of differential
	    operators, the Virasoro algebra and a q-analogue'},
	    \cmp {\bf 146}, 343 (1992)}
\ref\narganes{F. J. Narganes-Quijano, \jpa {\bf 24}, 593 (1991)}
\ref\woodhouse{N. Woodhouse, {\it `Geometric Quantisation'},
	 (Oxford University, 1980), p. 2}
\ref\caldi{A. Chodos and D. G. Caldi, {\it Quantum deformations
of the Heisenberg equations of motion'}, \jpa {\bf 24}, 5505 (1991)}
%
%

\Pubnum={QMW-92/19 \cr
	 hep-th/9211009}
\date={September 1992}
\pubtype={}

\nopagenumbers
\titlepage
\title{\bf A Quadratic Deformation of the Heisenberg-Weyl
and Quantum Oscillator Enveloping Algebras.}
\
\author{ Jens U. H. Petersen
		   \footnote{\circ}{e-mail: j.petersen@qmw.ac.uk
		   (decnet: 19678::petersen) \hfill\break
	    Work supported by a S.E.R.C. Research Studentship}}
\
\address{Theoretical Physics Group \break
	 Physics Department \break
	 Queen Mary and Westfield College \break
	 Mile End Road \break
	 London E1 4NS, UK}

\abstract
{A new 2-parameter quadratic deformation of the quantum oscillator algebra
and its 1-parameter deformed Heisenberg subalgebra are considered.
An infinite dimensional Fock module representation is presented
which at roots of unity contains null vectors and
so is reducible to a finite dimensional representation.
The cyclic, nilpotent and unitary representations are discussed.
Witten's deformation of $\sltwo$
and some deformed infinite dimensional algebras are constructed
 from the $1d$ Heisenberg algebra generators.
The deformation of the centreless
Virasoro algebra at roots of unity is mentioned.
Finally the $SL_q(2)$ symmetry of the deformed Heisenberg
algebra is explicitly constructed.}

\endpage

\pagenumber=1
\pagenumbers

\def\sect#1{\chapter{#1}}


\sect{Introduction}

In the recent intense study of quantum groups and
quantum (enveloping) algebras \cite{\jimbo,\drinfeld,\frtworon,\reviews}
the $q$-analogues of the simple Lie algebra $\sltwo$ and
of the non-semisimple quantum oscillator Lie algebra $\hfour$
have played an important r\^ole.
In this paper I consider a new
deformation of the quantum oscillator algebra $\hfour$,
which is nearer to the spirit of the Woronowicz program
in quantum groups than the deformations of $\hfour$ previously considered.

The deformations of $\sltwo$, $\slqtwo{q}$,
have been studied in detail: both the {\it transcendental}
deformations ($[X_+,X_-]=[N]_q$, $[N,X_\pm]=\pm X_\pm$)
of Jimbo\cite{\jimbo} and Drinfeld\cite{\drinfeld},
and the {\it quadratic} deformations of Sklyanin\cite{\sklyanin},
Woronowicz\cite{\woronoSU}, Witten\cite{\witten}
and Fairlie\cite{\fairlie} (see section 3).
When the deforming parameter is not a root of unity, these deformations have
a representation theory essentially equivalent to that of
the Lie algebra $\sltwo$ \cite{\rosso,\curtZach}.
The most interesting aspect of quantum algebras is when the
deformation parameter is a root of unity (see later),
then their properties are quite different from those of
the corresponding Lie algebras.

The Heisenberg algebra $\hthree$ and
the quantum oscillator algebra $\hfour$
are undoubtedly two of the most important
non-semisimple Lie algebras
in modern quantum physics.
Deformations of the universal enveloping algebra of the
quantum oscillator algebra have also been investigated
\cite{\macBieden-\CGST}.
Macfarlane and Biedenharn\cite{\macBieden} were the first to discuss the
non-linear $q$-deformation of the harmonic oscillator algebra
in the context of quantum groups, and from two independent
$q$-oscillators they realised a Jimbo-Drinfeld-deformation of $\sltwo$.
I call it the transcendental deformation of $\hfour$; it is
generated by $\setof{N,a_+,a_-}$, which satisfy the relations:
$$ [N,a_\pm]=\pm a_\pm\qquad a_-a_+ -q a_+ a_- = q^{-N}. \eqn\transOscill$$
Chan et al.\cite{\beijing}
studied the transcendentally deformed $su(2)$ algebra
(again using the Jordan-Schwinger construction)
both classically as a $q$-deformed
Poisson bracket algebra and in the quantum case
as a deformed Lie algebra, emphasising that deformation ($q$)
and quantisation ($\hbar$) are different concepts.
Yan\cite{\yan} presented the Hopf algebra structure of a
different transcendental deformation of the quantum oscillator algebra
(constructing its coproduct, coinverse, counit, and so on).
Celeghini et al.\cite{\CGST} have produced another simple deformation of
the quantum oscillator algebra as a quantum group with
the deformation in the Heisenberg subalgebra
($[a_-,a_+]=[e]_q$, $e$ being central).
Also Gelfand and Fairlie\cite{\fairlieGelfand} have studied
$q$-symmetrised polynomial algebras and their central extension
using a $q$-Heisenberg algebra.

In this paper, a 2-parameter quadratic deformation of
the quantum oscillator algebra $\hfour$
(and consequently also a 1-parameter deformation of its Heisenberg
subalgebra $\hthree$) is studied. In the next section the notion of
quadratic deformations of enveloping algebras is reviewed.
Then in section 3, the deformed quantum oscillator algebra
$\hqfour{q,r}$ and deformed Heisenberg subalgebra $\hqthree{r}$
are presented. Section 4 deals
with their representation theory, with emphasis on the possibility of
finite dimensional representations. In particular I discuss
the cyclic and `nilpotent' algebras and representations.
A unitary representation is also presented.
Section 5 contains the construction of some
algebras from $\hqthree{q}$ generators,
including a 1-parameter deformation of $\hfour$,
a quadratically deformed $su(1,1)$ algebra and some deformed infinite
dimensional algebras, including the $q$-Witt algebra.
Before concluding, I present an $SL_{s^2}(2)$  symmetry of $\hqthree{s}$.


\sect{Quadratic Deformations}

The $q$-analogues of the simple Lie algebras are all known,
though they are still being studied actively.
Much less is known about non-semisimple Lie algebras
and about their $q$-analogues (transcendental or
quadratic).

It seems that the transcendental quantum algebras are often
surprisingly easy to work with and tend to enjoy a pleasing Hopf algebra
structure. On the other hand from the viewpoint of non-commutative
geometry quadratically deformed quantum (Lie)
algebras are more natural\cite{\woronoDG,\manin,\wz}.
In classical differential geometry it is the commutative
$k$-algebra of smooth $k$-valued functions $C^\infty (M,k)$
on a smooth manifold $M$ that are of central importance.
$C^\infty (M,k)$ contains in particular the functions,
whose restriction to local open sets in $M$, gives
$M$ a local coordinate structure.
In non-commutative differential geometry, non-commutative
algebras take the r\^ole $C^\infty (M,k)$ had classically.

The simplest example is a (finite) $n$-dimensional vector space
$V$ over a field $k$ (of characteristic zero).
Let $\setof{x_i \mid i=1,\dots,n}$ be a basis of $V$
and let $V^*$ be the vector space dual to $V$.
A (dual) basis in $V^*$ forms a set of linear coordinates
on $V$. Smooth functions on $V$ can be represented as polynomials
in these linear coordinates with coefficients in $k$.
The set of all such polynomials has a natural structure
of a commutative algebra (the function algebra on $V$)
and is isomorphic to the symmetric algebra $S(V^*)$ over $V^*$.
I will briefly recall the construction of $S(V)$ from the
{\it tensor algebra} $T(V)$ of $V$ \cite{\tensorAlgebra}, before
turning to some more interesting examples.
In the basis given for $V$, $T(V)$ is isomorphic to the
free unital algebra of formal (non-commuting) polynomials in the
generating basis (denoted $\free{k}{x_1,\dots,x_n}$).
I will be using this isomorphism in what follows.
The (associative) tensor product in $T(V)$ will be denoted by `$\tensor$'.
In this section I am going to be interested in 2-sided
ideals $I$ of $T(V)$ and particularly in the quotient algebras
$T(V)/I$. (Remember that for an algebra $A$ with a linear subspace $R$,
$I:=A \cdot R \cdot A \subseteq A$ is in general a proper
2-sided ideal in $A$, and so $A/I$ is a quotient algebra.)
As a first example, consider the symmetrising ideal
$I_{\rm sym}\subset T(V)$:
$$ I_{sym}:= \ideal{v\tensor w -w\tensor v \mid v,w\in V}
	  :=T(V)\tensor\setof{v\tensor w -w\tensor v \mid v,w\in V}
					      \tensor T(V)\eqn\symIdeal$$
which is 2-sided by construction. I will often use
the notation $\ideal{\spacedcdot}$
to denote the {\it 2-sided ideal} (without unity)
in the appropriate tensor algebra {\it generated}
by the elements in the angled brackets. The quotient algebra
$T(V)/I_{\rm sym}$ defines the symmetric algebra on $V$:
$$ S(V):=T(V)/I_{\rm sym} \equiv T(V)/
	  \ideal{v\tensor w -w\tensor v\ \mid v,w\in V} \eqn\symAlg$$
Similarly $A(V):=T(V)/\ideal{a\tensor b + b\tensor a\mid a,b\in V}$
defines a realisation of the antisymmetrised
exterior (grassmann) algebra on $V$, which is $2^n$ dimensional.
 From here on I will take $k$ to be the complex numbers \complex.

Recall that a ($n$ dimensional) {\it quantum vector space}\cite{\manin,\wz}
has the relations $y_i y_j=q y_j y_i$,
(or $q^{-\half} y_i y_j -q^{\half} y_j y_i=0$) for $i<j$
($i,j=1,\dots,n\geq2$) between the elements of a basis
$\setof{y_a \mid a=1,\dots,n}$ of its non-commutative
coordinate-functions. So it is natural to think of
$$S_q(V):=T(V)/\ideal{y_i\tensor y_j -qy_j\tensor y_i \mid i<j} \eqn\qplane$$
or equivalently
$T(V)/\ideal{q^{-\half}y_i\tensor y_j-q^{\half}y_j\tensor y_i}$
as the non-commutative algebra of functions of a quantum vector space.
Although the 2-sided ideal is written in a coordinate
dependent way, the defining relations are in fact $GL_q(V)$-covariant.
The space $S_q(V)$ has an equivalent interpretation as the algebra
of $q$-symmetrised polynomials\cite{\fairlieGelfand}.
In the same way $Q_{\hbar}:=\cfree{x,p}/\ideal{xp-px-\hbar i}$ can
heuristically be thought of as the general
`quantum phase space of a particle in one dimension'
(with `non-commutative phase space coordinates' $x$ and $p$) and
$Q_{q,\hbar}:=\cfree{x,p}/\ideal{xp-qpx-\hbar i}$ as
a `deformed quantum phase space'.
Here again the $\cfree{\cdot}$ symbol means the associative, unital
\complex-algebra of formal (non-commuting) polynomials generated freely
by the elements inside its brackets.
As $\hbar\rightarrow 0$ they reduce to the symmetric algebra $S(V_{(2)})$
and the quantum vector space $S_q(V_{(2)})$ respectively
($V_{(2)}$ is a 2-dimensional vector space).
Note that this deformation $Q_{q,\hbar}$ of the
Heisenberg algebra $\hthree$ is slightly different from the
$\hqthree{q}$ that I consider in detail later.
(1-parameter deformations of bosonic and fermionic
quantum mechanical phase space and their symmetries
have been studied by Zumino\cite{\zumino} in the $R$-matrix formalism.)
Note the difference between quantisation deformation ($\hbar$) and
non-commutative deformations ($q$) \cite{\beijing}.

Universal enveloping algebras play a crucial r\^ole in Lie algebra and
Lie group theory, specially in their representation
theory\cite{\tensorAlgebra}.
The exponential map from the Lie algebra $g$
to the Lie group $G$ gives an (algebraic) embedding
$G \hookrightarrow U(g)$. The universal enveloping algebra $U(g)$
is the topological dual of the algebra of
continuous (representative) functions on $G$ ($fun(G)$).
The {\it universal enveloping algebra} $U(g)$ of a Lie algebra $g$
(with Lie bracket $[\cdot\thinspace\cdot]$) can be constructed
 from the tensor algebra of its underlying vector space:
$$U(g):=T(g)/\ideal{V \tensor W - W \tensor V
			      - [VW] \mid V,W\in g} \eqn\uea$$
so that the ideal $\ideal{\spacedcdot}$
gives $U(g)$ equivalence relations
of the form $V\cdot W-W\cdot V = [VW]$.
The definition of this (2-sided) ideal can also be written as
$\ideal{(id_2 -P -F)(V \tensor W) \mid V,W\in g}$,
with $id_2(a\tensor b):=a\tensor b$
(the identity operator, $id_2:g\tensor g\mapsto g\tensor g$),
$P(a\tensor b):= b\tensor a$ (the `flip operator',
$P:g\tensor g\mapsto g\tensor g$) and $F(V\tensor W):=[VW]$
(the structure tensor, $F:g\tensor g\mapsto g$).

In a basis $\setof{Y^i \mid i=1,\dots,n:=\dim(g)}$
of $g$, $T(g)$ can be identified with $\cfree{Y^1,\dots,Y^n}$
and the enveloping algebra is constructed as
$U(g):=T(g)/\ideal{Y^i\tensor Y^j-Y^j\tensor Y^i -{f^{ij}}_k Y^k \mid
					  i,j=1,\dots,n}$.
Using these structure constants of the undeformed Lie algebra $g$
in this basis ($[Y^iY^j]={f^{ij}}_kY^k$),
I can then construct a quadratic
$q$-analogue of its enveloping algebra
--{\it a quantum enveloping algebra}-- with
(non-zero) deformation parameters $q_{(i,j)}\in \complex^*$:
$$U_q(g):=T(g)/\ideal{q_{(i,j)}X^i\tensor X^j-q_{(j,i)}X^j\tensor X^i
			    -{f^{ij}}_k X^k \mid i,j=1,\dots,n}.\eqn\quea$$
Here the $\setof{X^a\mid a=1,\dots,\dim(g)}$
are the generators of $g$ considered as
a vector space and summation over the index $k$ (only) is understood.
Symmetry requires that $q_{(j,i)}=(q_{(i,j)})^{-1}$.

As usual $U_q(g) \rightarrow U(g)$
in the limit as all the deforming parameters $q_{(i,j)} \rightarrow 1$.
The reader may wonder how this relates to the theory
of non-commutative (covariant) differential geometry developed by
Woronowicz and others \cite{\woronoDG,\manin,\wz};
so I will just comment that a general quadratic deformation
of the above type \quea\ will not be bicovariant and may not even
be left covariant, though a number of important examples
are both `quadratic' and covariant \cite{\woronoSU,\castellani}.
Quadratic quantum algebras have the advantage that their representations
are rather easier to study.
(A typical left covariant quantum (Lie) algebra
has relations $(id\tensor id - R -C)(Z^i\tensor Z^j)=0$ on its
generators $Z^a$ (\ie $Z^iZ^j-{R^{ij}}_{kl}Z^kZ^l=C^{ij}_kZ^k$),
with $R:g\tensor g \to g\tensor g$
and $C:g\tensor g \to g$ satisfying the braid Yang-Baxter equation
($R_{12}R_{23}R_{12}=R_{23}R_{12}R_{23}$)
and the $R$-Jacobi identity respectively. The $C$ and $R$ tensors
must satisfy some additional relations to have bicovariance
\cite{\woronoDG}.
The undeformed case then corresponds to $R=P$ and $C=F$.)
Often it is desirable to endow the quadratic quantum algebra $U_q(g)$ with
a non-trivial (non-cocommutative) Hopf algebra structure,
but in general this is not possible
for arbitrary choices of $q_{(i,j)}$. When it is possible,
the particular $U_q(g)$ is called a ``quantum group''.
(From here on, the formal associative `$\tensor$'
product symbol of $T(g)$ will either be implicit
or denoted by `$\cdot$', and I will not distinguish the
product symbols of $T(g)$ and $U_q(g)$.)

For the quantum vector space \qplane\ the coordinate
function algebra $S(V)$ of a vector space $V$ was
deformed to a non-commutative algebra $S_q(V)$,
making the geometry `non-commutative'.
So in the same way the non-commutative differential
analogues of the Lie algebra of left-invariant vector
fields on a Lie group lose
the anti-commutativity of Lie brackets, as happens
for example in $U_q(g)$.
So relations of the form ``$qXY-q^{-1}YX=A$'' defining the
the quantum universal enveloping algebra
are more appropriate geometrically than the much used transcendental
deformations mentioned above ($[X,Y]=[A]_q$) which
have (rightly) received an enormous amount of attention
recently in algebraic contexts.
The ``$qXY-q^{-1}YX=A$'' form of deformation is
more symmetric than the ``$XY-qYX=A$'' form that is sometimes
considered. However unlike in the latter and transcendental
deformations, $q=0$ is not allowed.


\sect{The quadratically deformed quantum oscillator algebra}

The original (1-parameter) quadratic
deformation of the $\sltwo$ (complexified $su(2)$)
was invented by Woronowicz\cite{\woronoSU}.
A similar deformation was found by Witten to occur
in the context of vertex models\cite{\witten}
and generalised by Fairlie\cite{\fairlie,\curtZach} to a 2-parameter
deformation of the
universal enveloping algebra, denoted $\slqtwo{q,r}$.
It is generated freely by $\setof{W_+,W_-,W_0}$ respecting the
relations over \complex:
		$$\eqalign{  [W_0,W_+]_q&=W_+ \cr
			     [W_-,W_0]_q&=W_- \cr
	  {\rm and} \quad [W_+,W_-]_{1 \over r} &=W_0, \cr} \eqn\qsltwo$$
where the notation $[X,Y]_s:=sX\cdot Y-s^{-1}Y\cdot X$ is introduced;
$q,r,s \in \complex^*$ (the set of non-zero complex numbers).
For example the
first relation reads $qW_0\cdot W_+ -q^{-1}W_+\cdot W_0=W_+$.

So $\slqtwo{q,r}:=T(\sltwo)/I_{\sltwo,qr}$, where $I_{\sltwo,qr}$ is the
2-sided ideal in $T(\sltwo)$ generated by elements
corresponding to the relations in \qsltwo,
\ie $I_{\sltwo,qr}\equiv\ideal{q W_0 W_+ - q^{-1} W_+ W_0 -W_+ \comma
qW_-W_0 - q_{-1} W_0 W_- - W_- \comma r^{-1} W_+ W_-  -rW_- W_+ -W_0}$.
It seems that this 2-parameter deformation is only a Hopf algebra for
certain values of $q$ and $r$: when $q:=r^2$
or\cite{\witten} $r:=q^2$.
Woronowicz's deformation for real $q$
corresponds to a left-covariant differential
calculus on the quantum group $SU_q(2)$,
\ie the $q$-analogue of the Lie algebra of left-invariant
vector fields on SU(2)\cite{\woronoSU}.

The {\sl 1d quantum oscillator algebra} is a non-semisimple Lie algebra
$\hfour$ which has 4 generators ($n$, $a_+$, $a_-$ and $e$ which is
central). The Lie brackets are:
 $$\eqalign{[na_+]&=a_+\cr
	    [a_-n]&=a_-\cr
	    [a_-a_+]&=e\cr
   {\rm and}\quad [xe]&=0,\quad \forall x\in h(4).\cr}\eqn\qho.$$
I will now consider a deformation (similar to the one above)
of the {\sl 1d} quantum oscillator algebra:
a 2-parameter deformation of the $\hfour$
universal enveloping algebra, which I denote $\hqfour{q,r}$, generated
by $\setof{A_+, A_-, N, E}$ over \complex\ with the relations:
$$\eqalign{[N,A_+]_q&=A_+\cr
	   [A_-,N]_q&=A_-\cr
	   [A_-,A_+]_r&=E\cr
{\rm and} \qquad
	   [x,E]&=0\quad (\forall x\in \hqfour{q,r}).\cr}\eqn\qoscill$$
In other words $\hqfour{q,r}:=T(\hfour)/I_{\hfour,qr}$,
where corresponding to \qoscill:
$$I_{\hfour,qr}:=\ideal{[N,A_+]_q -A_+\comma [A_-,N]_q -A_-\comma
[A_-,A_+]_r -E\comma [x,E] \mid x\in T(\hfour)};$$
see \symIdeal\ for explanation of the $\ideal{\spacedcdot}$ notation.
In order to discuss the $1d$ deformed Heisenberg subalgebra, I also define:
$$\eqalign{I_{\hthree,r}&:= \ideal{[A_-,A_+]_r -E\comma
		      [x,E] \mid x\in T(\hthree)} \cr
	    &\equiv T(\hthree)\cdot \cfree{[A_-,A_+]_r -E\comma
       [x,E] \mid x\in T(\hthree)}_0 \cdot T(\hthree) \cr} \eqn\qheisen $$
an ideal (without unity) in $T(\hthree)$ (the 0 subscript denoting that `1'
is not a generator), so I can then construct the quotient algebra
$\hqthree{r}:=T(\hthree)/I_{\hthree,r}$.
Clearly $T(\hthree)\subset T(\hfour)$ and $I_{\hthree,r}\subset
I_{\hfour,qr} \subset T(\hfour)$, so $\hqthree{r}\subset \hqfour{q,r}$.
\hfill \break
\Remark $\hqfour{q,r}$ is a 2-parameter generalisation of
the transcendental oscillator algebra,
since $\hqfour{1,q^{-1}}$ is equivalent to \transOscill\ under
the identifications $E\equiv1$ and
$A_\pm\equiv a_\pm q^{\half N +{1 \over 4}}$.

It is easy to check that the quantum algebra relations \qoscill\
satisfy the {\it `braid-associativity consistency conditions'}\thinspace:
that is, it does not matter which way a cubic monomial $X_1X_2X_3$ in
the generators of the quantum algebra is re-ordered: the result
is the same answer either way. For example in $\hqfour{q,r}$ the `braidings'
$N A_+ A_- = r^{2} N A_- A_+ -rEN = q^2 r^2 A_- N A_+ - qr^2 A_-A_+ - rEN=
r^2 A_- A_+ N - rEN$ and
$ N A_+ A_- = q^{-2} A_+ N A_- + q^{-1} A_+ A_- = A_+ A_- N
= r^2 A_- A_+ N - rEN $ give the same expression.


\sect{Finite Representations}

The following identities can easily be proved by induction
for all positive integer $m$:
$$\eqalign{[N,{A_+}^m]_{q^m}&\equiv[m]_q {A_+}^m \cr
	[{A_-}^m,N]_{q^m}&\equiv[m]_q {A_-}^m \cr
    [A_-,{A_+}^m]_{r^m}&\equiv[m]_r E\cdot {A_+}^{m-1},\cr}\eqn\powerRelat$$
where I define the (usual) `$q$-number' to be $[m]_q:=q^m(m)_q$ and
$(m)_s:=\sum_{i=1}^{m} s^{1-2i}, \quad \forall m \in \positive$.
Any deformation parameter is always assumed to be non-zero complex,
unless reality is specified. Also $(0)_q:=0$, so $[0]_q\equiv0$.
$(m)_1 \equiv m \equiv [m]_1$ and $[1]_s\equiv 1 \equiv s(1)_s$.

As I am dealing with a deformation of the quantum oscillator
algebra, I want to start by considering a Fock-module
representation of $\hqfour{q,r}$. A linear representation of an algebra $A$
on a vector space $V$ is defined to be a homomorphic linear action
`$\cdot$' of the algebra on $V$
(\ie a homomorphism $A\rightarrow {\rm End}(V)$).
For the vector space of my representation I define
$V^{(j,c)}_{q,r}:=\bigoplus_{n=0}^\infty \complex u_n$, such that
$\setof{u_n \mid n\in\natural}$ is a basis of $V^{(j,c)}_{q,r}$
($\natural$ is the set of natural numbers)
and $u_0$ is a vacuum vector (lowest vector):
$$\eqalign{N\cdot u_0&=ju_0 \cr
	   A_-\cdot u_0&=0 \cr
	   E\cdot u_0&=cu_0. \cr}\eqn\lws $$
$j$ and $c\neq 0$ are numbers.
It is natural to associate $c$ with $\hbar$ (Planck's constant).
The following consistent left-action of $\hqfour{q,r}$ then
makes $V^{(j,c)}_{q,r}$ a left $\hqfour{q,r}$-module \powerRelat:
$$\eqalign{ A_+\cdot u_k &= u_{k+1} \cr
	    A_-\cdot u_k &= c(k)_r u_{k-1} \cr
	    N\cdot u_k &= (q^{-2k}j + (k)_q) u_k \cr
	E\cdot u_k &= cu_k \qquad  k\in\natural \cr}\eqn\rep$$
So $u_k= {A_+}^k\cdot u_0$.
The representations at different $j$ and different $c$
are generically inequivalent.
\Remark the defining relations \qoscill\ of
$\hqfour{q,r}$ ($\hqthree{r}$) are invariant under
the automorphism $A_+\mapsto -A_+$, $A_-\mapsto A_-$,
$N\mapsto N$ and $E\mapsto -E$, so
another representation exists which is inequivalent
to \rep:
$A_+\cdot u_k:= -u_{k+1}$, $A_-\cdot u_k:=c(k)_r u_{k-1}$
and $E\cdot u_k:= -cu_k$ ($N$ unchanged).
Matrix representations of $\hqfour{q,r}$ can also be constructed.
 From here on I will concentrate on the $j=0$ representations,
the natural vacuum representations,
defining $V_{q,r}^{(c)}:=V^{(j=0,c)}_{q,r}$.
Since $\hqthree{r}\subset\hqfour{q,r}$, the restriction of the
above representation of $\hqfour{q,r}$ to the subalgebra $\hqthree{r}$
induces a representation of it on the same space; I denote
the $\hqthree{r}$-submodule by $V_r^{(c)}\equiv \hqthree{r}\cdot u_0$.

For a simple Lie algebra $g$ and a generic value of $q$,
the centre of $U_q(g)$ is just generated by
the $q$-analogues of the (universal) Casimirs.
However for the deformation
parameter at a root of unity (denoted $q_p$), this is
no longer true and the representation theory changes
drastically\cite{\conciniKac}. Basically this is because the
centre of $U_{q_p}(g)$ is larger than that of $U_q(g)$.
This is also the case for $q$-analogues of non-semisimple
Lie algebras: the centres are also enlarged at roots of unity\cite{\sunGe}.

Specifically: when $r$ is a non-trivial $2n$-th root of unity $r_n$
($n>1$), \ie $(r_n)^{2n}=1$ and $r_n \neq 1$,
${A_+}^n$ and ${A_-}^n$ are additional generators of the centre
of $\hqthree{r_n}$, as can be seen from \powerRelat\
using $[n]_{r_n}\equiv0$.
If additionally $q$ is a non-trivial $2n$-th root of unity $q_n$
then ${A_+}^n$ and ${A_-}^n$ lie in the centre of $\hqfour{q_n,r_n}$.
\Remark when $q=r$ there exists a 2-Casimir of $\hqfour{q}$,
$K:=A_+\cdot A_- -E\cdot N$, which coincides in form with the Casimir
of $\hfour$.

In the following I want to concentrate on the case when $r$
is a {\it primitive} $2p$-th root of unity $r_p$ ($p>1$)
$$      r=r_p \qquad r_p:=e^{i\pi /p}    \eqn\rootsofunity$$
Then considering the representation of $\hqfour{q,r_p}$ on
$V_{q,r_p}^{(c)}$ (and at the same time the representation
of $\hqthree{r_p}$ on $V_{r_p}^{(c)}$), it is seen that
$u_p$ has become a {\it null vector} (or singular vector)
in the representation:
$$      A_-\cdot u_p = c(p)_{r_p} u_{p-1} \equiv 0.    \eqn\nullvect$$
since $(p)_{r_p}\equiv 0$. This can also be seen as
$A_-\cdot u_p=A_-\cdot{A_+}^p\cdot u_0= {A_+}^p\cdot A_- \cdot u_0\equiv 0$
\lws. In fact $\setof{u_{kp}\mid k\in \positive}$ are all null vectors.
Consequently the infinite dimensional representation on
$V_{q,r_p}^{(c)}$ ($V_{r_p}^{(c)}$) is reducible to a finite dimensional
one. Note however that these representations are not completely reducible.
If I define ${\cal B}_+:=\setof{{A_+}^n \mid n=1,\dots,\infty}$,
a (trivial) subalgebra of $\hqfour{q,r_p}$ ($\hqthree{r_p}$),
the set ${\cal B}_+\cdot u_p$ generates
a subspace of $V_{q,r_p}^{(c)}$ ($V_{r_p}^{(c)}$).
Then the space $T_q^{(p)}:=V_{q,r_p}^{(c)}/{\rm Span}({\cal B}_+\cdot u_p)$
($T^{(p)}:=V_{r_p}^{(c)}/{\rm Span}({\cal B}_+\cdot u_p)$) carries
a quotient module
representation of $\hqfour{q,r_p}$ ($\hqthree{r_p}$),
which is (finite) $p$-dimensional and irreducible.
In other words the quotient representation
is constructed by identifying $u_p\in V_{q,r_p}^{(c)}$
($V_{r_p}^{(c)}$) with $0$, $u_p=0$,
as is normally done with null vectors.
So $T_q^{(p)}=\sum_{n=0}^{p-1} \complex u_n\enspace=T^{(p)}$.
\Remark in this case the matrix representations mentioned earlier
can also be reduced to a finite $p\times p$ dimensional matrix
representation.

I will briefly discuss the reducibility of
these quotient representations at a general
(non-primitive) $2p$-th root of unity.
Then $r$ can take the values $r=(r_p)^n = e^{n\pi i/p}$ ($n=1,\dots,p-1$).
When $p$ is a prime number, the quotient representations are still
irreducible for all $n=1,\dots,p-1$.
If $p$ is not prime and $m:=\gcd(p,n)=1$, then the quotient
representations are also irreducible.
In the final case of $p$ not prime and $m=\gcd(p,n)>1$
the quotient representation is reducible
to a ${p \over m}$ dimensional representation of $\hqfour{q,{r_{p/m}}}$
($\hqthree{r_{p/m}}$), \ie $u_{p/m},u_{2p/m},\dots,u_{(m-1)p/m}$
are null vectors in $T_q^{(p)}$ ($T^{(p)}$).
Here the $\gcd(\cdot,\cdot)$ function takes the value of the greatest common
(integer) divisor of its arguments.

I will just mention that there is also a representation
of $\hqfour{q,r}$ ($\hqthree{r}$) on
$\bigoplus_{n\in\natural} \complex u_n$, that treats $A_+$ and $A_-$
symmetrically:
$$ A_+\cdot u_k=\left((k+1)_r\right)^\half u_{k+1} \quad
   A_-\cdot u_k=c\left((k)_r\right)^\half u_{k-1},    $$
($N$ and $E$ acting as in \rep).
For $r$ not at a (non-trivial) root of unity, it is equivalent
to \rep. But when $r=r_p$ for example,
$A_-\cdot u_{kp}\equiv0$ and $A_+\cdot u_{kp-1}\equiv 0$
(for all $k\in\positive$), and this representation {\it is} therefore
completely reducible to a $p$-dimensional one.

Since at $(q,r)=(q_p,q_p)$, ${A_+}^p$ and ${A_-}^p$ are
central in $\hqfour{q_p}:=\hqfour{q_p,q_p}$
($\hqthree{q_p}$), as I mentioned
earlier, it is natural to consider realisations with them
as complex numbers: say $\lambda:={A_+}^p$ and $\mu:={A_-}^p$.
Since I am concentrating on lowest vector representations,
I will just deal with $\lambda$, taking $\mu=0$.
${A_+}^p$ and ${A_-}^p$ generate a subalgebra of the centre
of $\hqfour{q_p}$ ($\hqthree{q_p}$), which is an ideal.
I define:
$$  \hqfour{q_p}^{(\lambda)}:=\hqfour{q_p}/\ideal{{A_+}^p -\lambda\comma
						{A_-}^p}  $$
(with a similar definition of $\hqthree{q_p}^{(\lambda)}$).
There are two particular cases I want to discuss:

\noindent
(i) In the case $\lambda\not=0$, I call the algebra
$\hqfour{q_p}^{(\lambda\not=0)}$
($\hqthree{q_p}^{(\lambda\not=0)}$) {\it cyclic} and
its representations correspond to the subset of
(lowest vector) representations
of $\hqfour{q_p}$ ($\hqthree{q_p}$) which are called cyclic\cite{\cyclic}.
For example a representation on the $p$-dimensional vector space
$T_q^{(p)}$ ($T^{(p)}$),
similar to above \rep\ except that $A_+$ acts as
$A_+\cdot u_{k}:= u_{(k+1)}$ ($k=0,1,\dots,p-2$) and
$A_+\cdot u_{p-1}:=\lambda u_0$, is cyclic.
In particular if $\lambda=1$ the action of the
subalgebra $\setof{{A_+}^k \mid k=1,\dots,p}$
on the cyclic representation space carries
a representation of the cyclic group ${\bf Z}_p$.
The cyclic representation space is obtained from
$V_{q_p,r_p}^{(c)}$ ($V_{r_p}^{(c)}$) with the
identification $u_k \equiv u_{k \thinspace {\rm mod}\thinspace p}$
($\forall k\in \natural$).
\Remark I should mention that with $\mu={A_-}^p\neq 0$
the fully cyclic representations can be constructed. They have neither
highest nor lowest vectors.

\noindent
(ii) The second case is $\lambda=0$. Then I call the algebra
$\hqfour{q_p}^{(0)}$ ($\hqthree{q_p}^{(0)}$)
{\it `nilpotent'}. The irreducible lowest vector representations
of the nilpotent algebras are also finite dimensional.
The finite $p$-dimensional representations of $\hqfour{q_p}$
($\hqthree{q_p}$) I mentioned above is an example of what I call
a nilpotent representation.
It would appear that $\hqthree{q_p}$ plays a significant r\^ole
in parafermionic quantum mechanics.
A nilpotent algebra very similar to
$\hqthree{q_p}^{(0)}$ has recently been discussed in the context
of paragrassmann algebras\cite{\filippov}.

In the case $q=1$ $\hqfour{1,r_p}$ has
the finite representation on $T_1^{(p)}$ described above.
This is possible since ${A_+}^p$ can still act nilpotently,
even though it is not in the centre of the algebra.
Then the $N$-eigenvalues become integer,
and it is meaningful to call $N$ the number operator:
$N\cdot u_k=(k)_1 u_k\equiv k u_k$ ($k=0,1,\dots,p-1$).
The representation of $\hqfour{1,r_p}$ on $T^{(p)}_{q=1}$
corresponds to a $p$-paragrassmann ($p$-parafermionic) oscillator
\cite{\beijing}:
$$[N,A_+]=A_+ \quad [A_-,N]=A_- \quad [A_-,A_+]_{r_p}=E \eqn\parafermion$$
Of course as $p \rightarrow \infty$, then $r_p \rightarrow 1$ and
the usual infinite dimensional bosonic Fock space representation
of $\hqfour{}$ is recovered.
The nilpotent algebra $\hqfour{r_p}^{(0)}$ is a $p$-paragrassmann algebra,
with $N$ interpreted as a $q$-number operator.

To conclude this section, I will discuss a (complex) unitary representation.
Note that I now take $q$ to be real and $r$ to be a positive real number.
As in the case of the transcendental oscillator \transOscill\ there exists
an anti-automorphism $\omega$ of $\hqfour{q,r}$ ($\hqthree{r}$) that maps
$\hqfour{q,r}\rightarrow \hqfour{q,r}$
($\hqthree{r}\rightarrow\hqthree{r}$):
$$\eqalign{   &A_+\mapsto A_- \qquad A_- \mapsto A_+ \cr
	      N \mapsto &N \quad E\mapsto E\comma \quad
	      \alpha \mapsto \alpha^*\enspace
	      \forall \alpha\in\complex \cr} \eqn\automor $$
preserving the defining relations \qoscill;
$\omega(x\cdot y)=\omega(y)\cdot\omega(x)$ ($\forall x,y \in \hqfour{q,r}$
($\hqthree{r}$)).
I define the following
positive definite sesquilinear scalar product $(\cdot,\cdot)$ on
the complex vector space $V^{(j,c)}_{q,r}$ ($V^{(c)}_r$):
$$     (u_k,u_l):= \delta_{k,l} \prod_{m=1}^k (m)_r
	  \equiv \delta_{k,l} (k)_r! \qquad k,l\in\natural \eqn\symmform $$
which being contravariant with respect to $\omega$
(\ie $(x\cdot u_k,u_l)=(u_k,\omega(x)\cdot u_l)$ and
$(u_k,y\cdot u_l)=(\omega(y)\cdot u_k,u_l)$
$\forall x,y\in \hqfour{q,r}$ ($\hqthree{r}$) and $k,l\in \natural$),
affords a unitary representation
of $\hqfour{q,r}$ ($\hqthree{r}$).
I define $(k)_r!:=(k)_r(k-1)_r\dots (1)_r$ ($k\in\positive$)
and $(0)_r!:=1$.
The basis can be normalised as
$u'_k:={1 \over \left((k)_r!\right)^\half} u_k$ ($k\in\natural$),
so that $(u'_k,u'_l)=\delta_{k,l}$.
There are two reasons why this unitary representation unfortunately cannot
be extended to the case at roots of unity: when $q$ and $r$ are
not real but treated as complex numbers
(i) the map $\omega$ is no longer an anti-automorphism and
(ii) the scalar product is no longer positive definitive:
$(u_k,u_k)=(k)_r!$ is not positive real.


\sect{Construction of other algebras}

In this section I will construct some well known
finite and infinite dimensional algebras from
the generators of $\hqthree{s}$, which I
denote in this section by $\setof{a_+,a_-,e}$
(not to be confused with \qho).
I will also briefly present the contraction of
$\slqtwo{1,r}$ to $\hqthree{r}$.
First I repeat the definition of $\hqthree{s}$ for completeness.

The defining relations which generate $\hqthree{s}$ are
$$\eqalign{  [a_-,a_+]_s&=e \cr
	     [e,a_\pm]&=0 \cr}. \eqn\qheisenberg $$
$\hqthree{s}:=T(\hthree)/I_{\hthree,s}$, where the ideal $I_{\hthree,s}$
defined earlier \qheisen, corresponds
to the relations of \qheisenberg.
In this section it will be useful to identify
$e$ with a scalar (\ie a multiple of unity) in
${U_s}^0\hthree = \complex$,
(as is normally done for central terms).
So I choose `$e=1$' and work with the realisation
$\hqthree{s}/\lbrace{e-1}\rbrace$.
Actually all the constructions can be made from
$\hqthree{s}$, but then factors of `$e$' appear regularly
on the right hand sides.

It is easy to check that $\setof{a_+,a_-,M}$ with
$M:= a_+ \cdot a_-$, satisfy
the generator relations isomorphic to those of $\hqfour{s}:=\hqfour{s,s}$:
$$\eqalign{  [M,a_+]_s&= a_+ \cr
	     [a_-,M]_s&= a_- \cr
	     [a_-,a_+]_s&=1(\equiv e), \cr} \eqn\bilinqoscill$$
so $\hqfour{s}$ is a subalgebra of $\hqthree{s}$.
\Remark it is equally good to define $M$ as
$\half(a_+\cdot a_- + a_-\cdot a_+ + \alpha e)$, $\alpha$ a complex number.
Next defining:
$$\eqalign{  &B_+:= a_+ \cdot a_+ \qquad B_-:= a_- \cdot a_- \cr
	     &B_0:= \half (s^2 a_- \cdot a_+ +s^{-2}a_+ \cdot a_-) \cr}
					\eqn\BgensDef$$
I obtain the following realisation of a deformation of $su(1,1)$,
provided $s\not\in\setof{e^{i\pi \over 4},e^{2\pi i \over 4},
e^{3\pi i\over 4}}$:
$$\eqalign{  [B_0,B_+]_{s^2} = \half [2]_{s^2}& [2]_s B_+ \qquad
	     [B_-,B_0]_{s^2} = \half [2]_{s^2} [2]_s B_- \cr
	     &[B_-,B_+]_{s^4} = 2 [2]_s B_0 \cr} \eqn\Brelat$$
and so $U_{s^2} su(1,1)\subset\hqthree{s}$.
If I define:
$$     W_0 :=  2 \left( [2]_{s^2} [2]_s \right)^{-1} B_0 \qquad
       W_{\pm} :=\pm \left(([2]_{{s^2}})^\half [2]_s \right)^{-1} B_{\pm}
						 \eqn\wwgens$$
then it is easy to check that $\setof{W_0, W_+, W_-}$ satisfy
the defining relations \qsltwo\ of Witten's
deformation of $su(2)$,
$\slqtwo{s^2}\equiv\slqtwo{{q=s^2},{r=s^4}}$,
contained in $\hqthree{s}$ as a subalgebra.
Since $\sltwo\simeq sp_2$, I am also free to call this a deformation
of the Lie algebra $sp_2$.

Next I mention that elements $\setof{a_-,{a_+}^k \mid k\in\natural}$
of $\hqthree{s}$ generate a subalgebra
with $[a_-,{a_+}^k]_{s^k}=[k]_s {a_+}^{k-1}$.
This subalgebra has a natural interpretation as
the algebra of polynomials in one variable with a $q$-derivation and
then $\hqthree{s}$ is the algebra of polynomials and q-differential
operators.
When $s=s_p$ \rootsofunity, then repeating the
methods of section 4, the cyclic and nilpotent forms
of this algebras can be studied.

To my knowledge, contractions of quantum groups
were first performed in ref. \celeghini, where the transcendental
deformation of $\sltwo$ was contracted to the Heisenberg and
Euclidean algebras.
Here I show that this is also possible with
the ($q=1$) quadratic deformation of $\sltwo$, $\slqtwo{1,r}$
\qsltwo.
I scale the generators: $W_0\rightarrow X_0:=\xi W_0$
and $W_\pm \rightarrow X_\pm:=\xi^\half W_\pm$ $(\xi \geq 0)$. Then in the
limit $\xi \rightarrow 0$, I find that the new `contracted' algebra
has $X_0$ central:
$$\eqalign{       [X_\pm,X_0] &= 0 \cr
		  [X_+,X_-]_r &= X_0. \cr}     \eqn\contraction$$
These are the defining relations of $\hqthree{r}$. So I have
contracted $\slqtwo{1,r}$ to $\hqthree{r}$. I can also
scale the generators like this: $W_\pm \rightarrow \xi^{-1} W_\pm$
and $W_0 \rightarrow W_0$, and in the limit as $\xi\rightarrow 0$
the contracted algebra becomes a quadratic deformation of the
Euclidean algebra: $[W_0,W_+]_s=W_+$, $[W_-,W_0]_s=W_-$
and $[W_+,W_-]=0$.

\medskip
\def\vir{$Witt$}

Recently Kassel has found a $q$-analogue of the Virasoro
algebra 2-cocycle\cite{\kassel}, \ie a general central extension
for the $q$-Virasoro algebra (he also briefly discusses
the $q$-Heisenberg relation). Here I construct
a $q$-deformation of the Witt algebra \vir\
(the (centreless Virasoro) Lie algebra of vector fields on the circle).
It is simpler than that of ref. \chaiKulLuk\ which
involves the $N$ of the transcendental $q$-oscillator algebra \transOscill.
My deformation is similar to the centreless one in ref. \kassel.
Incidently the generators of the $q$-\vir\ algebra in ref. \curtZach\
appear in my notation as simply $Z_m=x^m(x\partial)_{1 \over r}$, with
$[Z_m,Z_n]_{r^{n-m}}=[m-n]_rZ_{m+n}$.

The set $\setof{L_{-1}:=a_-\comma L_0:=a_+ \cdot a_-\comma
L_1:=(a_+)^2\cdot a_-}$ generates a deformation of
the usual $su(1,1)$ subalgebra of \vir:
$$ [L_1,L_0]_{s^{-1}}= L_0 \qquad [L_1,L_{-1}]_{s^{-2}}=[2]_s L_0
    \qquad [L_0,L_{-1}]_{s^{-1}}= L_{-1}$$
The set $\setof{L_m:=-(a_+)^{1+m}\cdot a_- \mid m=-1,0,1,2,\dots}$
generates a deformation of the `constraint subalgebra'
of the full $s$-Virasoro algebra, which I construct next.
I formally extend $\hqthree{s}$ to an algebra ($\hqthree{s}'$)
additionally generated by ${a_+}^{-1}$, with the extra relations:
$$\eqalign{     {a_+}^{-1} \cdot a_+ &=1=a_+ \cdot {a_+}^{-1} \cr
		[{a_+}^{-1},a_-]_s&=e \cdot {a_+}^{-1} \cdot {a_+}^{-1} \cr
		[{a_+}^{-1},e]&=0. \cr }  \eqn\aMinus$$
then $L_m:=-(a_+)^{1+m} \cdot a_-$ ($m \in \integer$) generates
a q-deformation of \vir\ realised in $\hqthree{s}'$:
$$       [L_m,L_n]_{s^{n-m}} = [m-n]_s L_{m+n}.   \eqn\qVir$$
Here $(k)_q$ and $[k]_q$ are extended to non-positive
integers: $(-k)_q=-q^{-2k}(k)_q$ and $[-k]_q=-[k]_q$,
$\forall k\in \positive$.
\Remark it is also possible to construct a centreless
$q$-deformed $w_\infty$ algebra\cite{\chaiKulLuk}
($W^{(k+1)}_n:= {a_+}^{n+k} \cdot {a_-}^k$ $k\in\positive$),
as was done in ref. \fairlieGelfand\ using a $q$-Heisenberg algebra.

Considering this construction of $s_p$-\vir\ instead with the cyclic
algebra of $\hqthree{s}^{(\lambda=1)}$,
then the $s_p$-\vir\ algebra has the relations:
$L_{m+kp}^{cycl}\equiv L_m^{cycl}$ ($k\in \integer$ and $m\in\natural$),
and $L_{-m}:=L_{p-m}$ ($m\in\natural$).
So this deformation of $s_p$-\vir\ realised in $\hqthree{s}$ has
$p$ generators $\setof{L_0^{\rm cycl}\comma L_1^{\rm cycl},\dots,
L_{p-1}^{\rm cycl}}$. It might be interesting to study this
algebra in more detail.
{\sl Note added:\ }after finishing this work I became
aware that cyclic representations of a different cyclic $q$-\vir\
algebra have been considered previously\cite{\narganes}.


\sect{The Symmetry of $\hqthree{s}$}

Consider the matrix:
$$ T:=\left( \matrix{a & b \cr
		  c & d \cr} \right)$$
with the following relations between its non-commuting
elements ($q \in\complex^*$, a non-zero complex number):
$$  \eqalign{ ac=qca \qquad & ad-qcb=1 \cr
	      bd=qdb \qquad & da-q^{-1}bc=1 \cr }  \eqn\SpqRelat$$
which are in fact a deformation of the defining relations
of the symplectic matrix group $Sp(2n)$ in the case n=1 \cite{\woodhouse}.
I call this quantum algebra
$Sp_q'(2):= \cfree{a,b,c,d}/
\ideal{ac-qca\comma bd-qdb\comma ad-qcb-1\comma da-q^{-1}bc-1}$.
It contains the well-studied quantum group $SL_q(2)$ as
a subalgebra, since $SL_q(2)$'s defining relations
include \SpqRelat. Its additional relations are:
$$      bc=cb \quad ab=qba \quad cd=qdc.        \eqn\extraSLq $$
It is easy to show that $Sp_{s^2}'(2)$ is a symmetry of
the `deformed quantum phase space' $\hqthree{s}$:
the form of the defining relation $[A_-,A_+]_s=E$
is left-covariant with respect to the left co-action
($\hqthree{s}\rightarrow Sp_{s^2}'(2) \tensor \hqthree{s}$):
$$     \left(\matrix{ A_+ \cr A_- \cr} \right) \mapsto
       \left(\matrix{ A_+' \cr A_-' \cr} \right):=
       T\dtensor \left(\matrix{ A_+ \cr A_- \cr} \right)
		  \quad E \mapsto E':=1\tensor E   \eqn\leftAction$$
\ie $[A_-',A_+']_s\equiv
[c\tensor A_+ + d\tensor A_-, a\tensor A_+ + b\tensor A_-]_s\equiv E'$
follows using \SpqRelat. More precisely under the left co-action:
$\hqthree{s}\equiv \cfree{A_+,A_-,E}/
\ideal{[A_-,A_+]_s-E\comma [E,x] \mid \forall x} \rightarrow
\cfree{A_+',A_-',E'}/
\ideal{[A_-',A_+']_s-E'\comma [E',x] \mid \forall x}
\simeq \hqthree{s}$.

On the other hand it is equally good to consider the right co-action
($\hqthree{s}\rightarrow \hqthree{s} \tensor Sp_{s^2}''(2)$):
$$     \left( A_+ \enspace  A_-  \right) \mapsto
       \left( A_+'' \enspace A_-'' \right):=
       \left( A_+ \enspace A_- \right) \dtensor T
		   \quad E\mapsto E\tensor 1      \eqn\rightAction$$
and this results in the relations ($q=s^2$):
$$  \eqalign{ ab=qba \qquad & ad-qbc=1 \cr
	      cd=qdc \qquad & da-q^{-1}cb=1 \cr}  \eqn\rightRelat$$
which is another deformation of the $Sp(2)$ defining relations.
So requiring both left and right covariance (bicovariance) of
the $\hqthree{s}$ symmetry, means combining the relations
\SpqRelat\ and \rightRelat, forcing the extra relation $bc=cb$.
Then it is seen that the bicovariant symmetry of $\hqthree{s}$
is the quantum group $SL_{s^2}(2)$.
 From the definitions of the left $SL_{s^2}$--co-action
$\Delta_L$ \leftAction\ and the right co-action $\Delta_R$ \rightAction,
it is almost obvious that they are compatible with
the $SL_{s^2}$-coproduct ($\Delta(T):=T\dtensor T$),
\ie that $(\Delta_L \tensor id) \circ \Delta_L=
(id \tensor \Delta)\circ \Delta_L$ and
$(id \tensor \Delta_R) \circ \Delta_R=
(\Delta \tensor id)\circ \Delta_R$, and
also that they co-commute: $(id \tensor \Delta_R)\circ \Delta_L=
(\Delta_L \tensor id)\circ \Delta_R$.
I will not consider the full linear quantum symmetry
of $\hqfour{q,r}$ here, though I comment that this
$SL(2)_{s^2}$ symmetry can be extended to the realisation
of $\hqfour{s}$ in $\hqthree{s}$ constructed earlier \bilinqoscill.

Finally I want to make a remark about $SL_q(2)$ when $q$ is
a non-trivial $p$-th root of unity $q_{(p)}$.
Then it turns out that $b^p$ and $c^p$
lie in the centre of the quantum group algebra, whereas
$a^p$ and $d^p$ only commute with polynomials in $b$ and $c$.
This is significantly similar to the situation
discussed in section 4, where ${A_+}^p$ and ${A_-}^p$
fell in the centre of $\hqfour{q_p}$, but $N^p$ did not,
and corresponds in fact to the centrality of ${X_+}^p$
and ${X_-}^p$ in the transcendental $\slqtwo{q_{(p)}}$.


\sect{Conclusions}

It may be that only for $\sltwo$ do we have the equivalent
Hopf algebras of Drinfeld-Jimbo, Woronowicz and Witten.
The relationship between transcendental and quadratic
quantum algebras still requires more study.
I have tried to construct a coproduct
for $\hqfour{q,r}$, but this does not seem to be possible.
A more complicated quadratic deformation of $\hthree$ and $\hfour$,
such as a general `bicovariant q-Lie algebra',
may give a nice Hopf algebra structure.
Though I would comment that $\hqfour{q,r}$
($\hqthree{r}$) does not really need to be a Hopf algebra,
since it is only really the symmetries of the system that
are expected to be quantum groups.

There are still some issues which I have not discussed in this paper.
One of these is the $q$-quantum mechanics\cite{\caldi} of $\hqfour{q,r}$
($\hqthree{r}$) when $q$ and $r$ are complex, in particular when they
are roots of unity. If it were possible to find a unitary
representation in this case, then
the Hilbert space of $\hqfour{q,r_p}$ ($\hqthree{r_p}$)
would be finite dimensional and it would be interesting
to study the cyclic and nilpotent representations.
At present unitary vector space representations of the
transcendental \transOscill\ and quadratic \qoscill\ oscillators
are only known for positive real deformation parameters.
It may be possible to consider representations on quantum spaces,
where a generalisation of unitarity may exist.

Work is now in progress on $q$-deformed affine algebras,
moving from this quantum mechanical algebraic framework into
the quantum field theory arena.


\vskip 0.5truein

\ack{
I would like to thank Chris Hull for encouraging me to work on
quantum groups, specially the $q$-quantum oscillator, for much
helpful advice and for carefully reading through drafts of this paper.
Also thanks to Paul Wai for useful discussions
and to David Fairlie and Cosmas Zachos for helpful comments.
My S.E.R.C. Research Studentship,
without which this work could not have been undertaken,
is gratefully acknowledged.}

\refout

\bye